\documentclass[12pt]{article}

\usepackage{setspace}
\usepackage[english]{babel}
\usepackage{pdfpages}
\usepackage[letterpaper,top=2cm,bottom=2cm,left=1in,right=1in,marginparwidth=1in]{geometry}
\usepackage[mathscr]{euscript}
\usepackage{colortbl}
\usepackage[table]{xcolor}
\usepackage{subcaption}
\usepackage{amsmath}
\usepackage{amsthm}
\usepackage{verbatim}
\usepackage{amsfonts}
\usepackage{graphicx}
\usepackage{enumitem}
\usepackage{booktabs, soul}
\usepackage[round,authoryear]{natbib}
\usepackage{algorithm}
\usepackage{algorithmicx}
\usepackage{algpseudocode}
\usepackage{threeparttable}
\usepackage{multirow}
\usepackage[colorlinks=true, allcolors=blue]{hyperref}

\newcommand{\on}{\cellcolor{gray!25}{\rule{0pt}{1.2em}\centering$\Box$}}
\newcommand{\off}{\phantom{\cellcolor{white!25}{\rule{0pt}{1.2em}\centering$\Box$}}}

\title{BOIN Designs for Dose Escalation With Selected Dose Combinations in Oncology Phase I Trials}
\author{Yuxuan Chen$^1$, Haiming Zhou$^1$, Keiko Nakajima$^2$, Philip He$^{*1,3}$}

\begin{document}

\maketitle

\section*{Author Affiliations}

\noindent
$^1$Biostatistics, Daiichi Sankyo, Inc., 211 Mt. Airy Rd, Basking Ridge, New Jersey, USA

\noindent
$^2$Clinical Development, Daiichi Sankyo, Inc., 211 Mt. Airy Rd, Basking Ridge, New Jersey, USA

\noindent
$^3$Rutgers University, Department of Statistics, 110 Frelinghuysen Rd, Piscataway, New Jersey, USA

\vspace{1in}
\noindent
$^*$Correspondence: Philip He, Ph.D.\\
Email: Philip.He@rutgers.edu

\newpage

\section*{Data Availability Statement}
No human participant data were used in this work. All results are reproducible from simulation, and the code used to implement the proposed methods, is available at \url{https://github.com/chnyuxuan/BOIN-CX}.

\section*{Funding Statement}
This research received no specific grant from any funding agency in the public, commercial, or not-for-profit sectors.

\section*{Conflict of Interest Disclosure}
Y.C., H.Z., K.N., and P.H. are employees of Daiichi Sankyo, Inc. and may own its stocks.

\section*{Disclaimer}
Please note that the views and opinions expressed in this paper are those of the authors and are not intended to reflect the views and/or opinions of their employer(s).

\newpage

\begin{abstract}
In phase I dose escalation studies for dual-agent combinations, at least one drug often has an established monotherapy dose. Consequently, substantial prior clinical safety data often exist for one or more monotherapies, allowing the study to focus on a subset of selected dose combinations rather than exhaustively evaluating all possible dose combinations for two agents. The Bayesian Optimal Interval (BOIN) design framework is widely recognized for its robust performance and ease of implementation; however, the BOIN for combination design, abbreviated as BOIN-C in this paper, was originally developed to evaluate full combinations and may not be directly applicable for the subset of selected combinations. In this paper, we propose three extensions to the BOIN-C design to address scenarios involving selected dose combinations: (a) BOIN-CS: a generalized BOIN-C design to accommodate any subset of dose combinations. (b) BOIN-CE: Exploration of new off-diagonal dose combinations when de-escalating. This option provides additional opportunities to treat patients with dose combinations that have not been administered.
(c)  BOIN-CB: Bayesian logistic regression model (BLRM)-guided BOIN design, which uses the BLRM model to break the tie when two dose combinations have an equal posterior probability of being selected. This can be useful when the dose-toxicity relationship is expected to be reasonably aligned with a logistic relationship. 
These study design options are motivated by practical considerations, and their operating characteristics are evaluated through extensive simulations under various scenarios, demonstrating satisfactory performance.
\end{abstract}

\newpage

\section{Introduction}\label{sec.intro}

Combination therapy is an important strategy in oncology drug development, because different agents can target complementary biologic pathways, improve depth or durability of response, and potentially delay therapeutic resistance. Between January 2011 and December 2021, 122 of 385 cancer therapy approvals are combination regimens \citep{fudio2022anti}. In early phase oncology drug development, dose finding usually consists of two stages: a dose-escalation phase to determine the maximum tolerated dose (MTD) and identify admissible doses, followed by a dose-optimization phase that compares multiple admissible doses \citep{he2025onco}.
In this paper, we discuss the dose escalation phase for combination therapy based on toxicity.

% The statistical methods for dose escalation design in monotherapy —such as the continual reassessment method (CRM) \cite{OQuigley1990}, modified toxicity probability interval 2 (mTPI-2) \cite{Guo2017}, Bayesian optimal interval (BOIN) design \cite{Liu2015}, and Bayesian logistic regression model (BLRM) with escalation with overdose control (EWOC) \cite{Tighiouart2010}—assume a unidirectional monotonic dose-toxicity relationship. These methods are generally not applicable to combination dose escalation design when multiple dose levels are evaluated for both drugs in a two dimensional dose space. A comprehensive discussion on design and conduct considerations for first-in-human trials is provided by \citep{Shen2019}.

In monotherapy dose-escalation studies, the dose–toxicity relationship is one-dimensional, and toxicity is typically assumed to increase monotonically with dose. Common designs used in this setting include the continual reassessment method (CRM) \citep{OQuigley1990}, the modified toxicity probability interval 2 (mTPI-2) \citep{Guo2017}, the Bayesian optimal interval (BOIN) design \citep{Liu2015}, and the Bayesian logistic regression model (BLRM) with overdose control (EWOC) \citep{Tighiouart2010}. In contrast, for dual-agent dose escalation studies, the admissible space becomes two-dimensional, making dose finding for combination therapies more complex and challenging, since escalation or de-escalation can proceed along multiple possible directions.

Multiple phase I dose escalation designs for combination therapies have been proposed to address the challenges posed by the complex, two-dimensional dose-toxicity relationships. These designs can be grouped into three categories: rule-based, model-based, and model-assisted designs.
Rule-based designs rely on predetermined escalation and de-escalation rules based on the number of observed dose-limiting toxicities (DLTs) among a fixed-size cohort. This approach does not assume any statistical model. A representative design is the $2 + 1 + 3$ design \citep{Fan2009}, which provides an alternative to the traditional $3 + 3$ design with defined rules for maximum tolerated dose combination (MTC) identification. Closely related nonparametric approaches also avoid parametric dose-toxicity modeling. For example, \cite{Ivanova2004} proposed a two-dimensional extension of the Narayana up-and-down design, and used bivariate isotonic regression to estimate the set of MTCs. While rule-based designs are simple to implement in clinical practice, they generally exhibit inferior operating characteristics compared to model-based approaches \citep{Ji2010} and are less flexible with respect to cohort size and target toxicity level. Model-based approaches, in contrast, specify a parametric model for the joint dose–toxicity relationship. Representative examples include the Bayesian design of \cite{Wang2005}, which uses a parsimonious working model for two-dimensional dose finding; the BLRM combination design \citep{neuenschwander2015bayesian}; the copula-type Bayesian adaptive design that captures drug synergy \citep{yin2009bayesian}; and the Bayesian logistic model treating one drug's dose level as a covariate \citep{Bailey2009}. Despite their flexibility and improved accuracy, model-based designs are often computationally intensive, requiring repeated model fitting after each cohort and relying on potentially unjustified parametric assumptions.

Model-assisted designs combine the robustness of model-based approaches with the simplicity of rule-based implementation. These methods derive dose-assignment rules from underlying statistical models but apply them using pre-tabulated, easy-to-follow rules during the trial. Examples include the product of independent beta probabilities design (PIPE) \citep{mander2015product}, the Bayesian Optimal Interval design for combinations (BOIN-C) \citep{lin2017bayesian}, the drug-combination Keyboard design (Keyboard-C) \citep{pan2020keyboard}, and the surface-free design \citep{mozgunov2020surface}. However, these methods cannot be directly applied to the setting that only a subset of dose combinations is considered and their use in practice is limited. 

The BOIN design has been widely adopted in early-phase oncology trials for monotherapy dose finding with favorable operating characteristics and ease of implementation \citep{Ananthakrishnan2022}. Its extension to dual-agent settings, the BOIN-C design \citep{lin2017bayesian}, has demonstrated a well-balanced performance relative to other combination designs, achieving relatively high correct selection rates across a range of scenarios \citep{barnett2024comparison}. However, BOIN-C was developed to evaluate full combination matrices and is not directly applicable when only a subset of dose combinations is selected. In practice, at least one agent often has an established monotherapy dose before the initiation of a combination dose escalation study. This existing clinical information can guide the choice of a subset of dose combinations, allowing for the efficient determination of the MTC or the recommended phase 2 dose combination (RP2DC).
 Table \ref{tab:setup} illustrates several examples of selected dose combinations included in a dose escalation study. In the special cases in scenarios Table \ref{tab:setup:a} and Table \ref{tab:setup:b}, the selected dose combinations are constructed to be monotonically increasing in dose levels, either by fixing one drug’s dose or by collapsing the two-dimensional combination space into a single dimension. For such cases, monotherapy dose escalation designs are appropriate \citep{fudio2022anti}. Scenario Table \ref{tab:setup:c} of complete dose combinations is not commonly used in practice. Scenario Table \ref{tab:setup:d} represents a subset of dose combinations, which motivates the development of the methodology presented in this paper. 
 
\begin{table}[htbp]
\centering

%--- helper for column header ------------------------------------------
\newcommand{\GridHeader}{%
  & \multicolumn{4}{c}{\textbf{drug B (mg)}} \\ \cmidrule(lr){2-5}
  \textbf{drug A (mg)} & 120 & 160 & 200 & 240 \\ \midrule
}

%--- helper macros for “on/off” cells (define as you like) --------------
% \newcommand{\on}{\cellcolor{gray!30}}
% \newcommand{\off}{}
\begin{minipage}[t]{0.48\textwidth}
  \centering
  \subcaption{$\mathcal{C}_{\text{mono1}}$: monotonic escalation matrix}
  \label{tab:setup:a}
  \begin{tabular}{rcccc}
    \toprule
    \GridHeader
     15 & \off  & \off  & \off & \off \\
     25 & \off & \off  & \off  & \off \\
     50 & \on & \on & \on  & \on  \\
     75 & \off & \off & \off  & \off  \\ \bottomrule
  \end{tabular}
\end{minipage}
\hfill % horizontal space between the two tables
\begin{minipage}[t]{0.48\textwidth}
  \centering
  \subcaption{$\mathcal{C}_{\text{mono2}}$: monotonic escalation matrix}
  \label{tab:setup:b}
  \begin{tabular}{rcccc}
    \toprule
    \GridHeader
     15 & \on  & \off & \off  & \off \\
     25 & \on  & \on  & \off & \off  \\
     50 & \off & \on  & \on  & \off \\
     75 & \off  & \off & \on  & \on  \\ \bottomrule
  \end{tabular}
\end{minipage}
%======================== 2nd ROW: single matrix ======================
\vspace{1.2em}  % vertical gap between rows
\par
\centering
\begin{minipage}[t]{0.48\textwidth}
  \centering
  \subcaption{$\mathcal{C}_{\text{Full}}$: complete $4\times4$ matrix}
  \label{tab:setup:c}
  \begin{tabular}{rcccc}
    \toprule
    \GridHeader
     15 & \on  & \on  & \on  & \on  \\
     25 & \on  & \on  & \on  & \on  \\
     50 & \on  & \on  & \on  & \on  \\
     75 & \on  & \on  & \on  & \on  \\ \bottomrule
  \end{tabular}
\end{minipage}
\hfill % horizontal space between the two tables
\begin{minipage}[t]{0.48\textwidth}
  \centering
  \subcaption{$\mathcal{C}_{s}$: prespecified partial matrix}
  \label{tab:setup:d}
  \begin{tabular}{rcccc}
    \toprule
    \GridHeader
     15 & \on  & \on  & \off & \off \\
     25 & \off & \on  & \on  & \off \\
     50 & \off & \off & \on  & \on  \\
     75 & \off & \off & \on  & \on  \\ \bottomrule
  \end{tabular}
\end{minipage}
\caption{Illustrative scenarios of dose combination spaces. Gray highlighted dose combinations are those included in the trial. Panels (a) and (b) are monotone paths, panel (c) is the full set of dose combinations, and panel (d) only contains a subset of dose combinations.}
\label{tab:setup}
\end{table}

We propose the BOIN-CS design, a practical extension of BOIN-C that enables the generalization of escalation and de-escalation to any prespecified subset of dose combinations. BOIN-CS preserves the features of the BOIN-C design—simple interval decisions, overdose control, and favorable operating characteristics—while allowing dose finding to be restricted to a smaller, more desirable subset of dose combinations. We also describe two other variants. BOIN-CE allows exploratory de-escalation to an adjacent, previously untried dose combination. BOIN-CB uses a Bayesian logistic regression model (BLRM) only for tie-breaking. The BLRM model incorporates existing data and guides the tie-breaking rather than random selection, as in BOIN-CS. 

The remainder of the paper is organized as follows. Section~\ref{sec.methods} introduces BOIN-CS and the two optional extensions. Section~\ref{sec.simulation} evaluates their operating characteristics in a comprehensive simulation study. Section~\ref{sec.case} presents a case study. Section~\ref{sec.discussion} concludes with practical guidance for implementation.

\section{BOIN-C design and its extensions}
\label{sec.methods}

Within the BOIN design framework \citep{Liu2015}, consider a phase I oncology dose-escalation study for a combination therapy consisting of two drugs, A and B. Let $d_{ij} = (d_i^A, d_j^B)$ denote the dose combination of drug A dose level $i$ and drug B dose level $j$, for $i = 1, \dots, I$ and $j = 1, \dots, J$. Let $\mathcal{C}_{Full} = \{d_{ij} : i=1,\dots,I;\, j=1,\dots,J\}$ be the full set of all dose combinations. Let $\pi_{ij}$ denote the probability of dose-limiting toxicity (DLT) at combination $d_{ij}$, and let $(n_{ij}, y_{ij})$ denote the number of treated patients and the number of observed DLTs, respectively.  The DLT rate is estimated as $\hat{\pi}_{ij} = y_{ij}/n_{ij}$. In the BOIN design framework, dose assignment is driven by a target toxicity probability $\phi$, an under-dosing threshold $\phi_1$, and an overdosing threshold $\phi_2$. The BOIN algorithm escalates the dose when $\hat{\pi}_{ij} \le \lambda_e$, de-escalates the dose when $\hat{\pi}_{ij} > \lambda_d$, and otherwise remains at the current dose, where the boundaries $\lambda_e$ and $\lambda_d$ are chosen to minimize the probability of incorrect escalation and de-escalation decisions locally, where 
\[
\lambda_e = \frac{\log\left(\frac{1 - \phi_1}{1 - \phi}\right)}{\log\left(\frac{\phi(1 - \phi_1)}{\phi_1(1 - \phi)}\right)}
\qquad \text{and} \qquad
\lambda_d = \frac{\log\left(\frac{1 - \phi}{1 - \phi_2}\right)}{\log\left(\frac{\phi_2(1 - \phi)}{\phi(1 - \phi_2)}\right)}.
\]

\subsection{BOIN-C design for the full set of dose combinations $\mathcal{C}_{Full}$}
\label{sec.BOIN.C}

The BOIN-C design \citep{lin2017bayesian} extends the BOIN design to dose escalation in combinations. Similar to the BOIN design, the BOIN-C design assigns the dose combination to the next cohort that has the largest posterior probability $\hat{\pi}_{ij}^{\mathrm{BOIN}}$ of falling in the optimal interval $(\lambda_e, \lambda_d)$, where
$\hat{\pi}_{ij}^{\mathrm{BOIN}} = P\!\left\{\pi_{ij} \in (\lambda_e, \lambda_d) \mid n_{ij}, y_{ij}\right\}.
$

If no patients have been assigned to the candidate dose combination, the probability is determined by the prior distribution alone. When the probability is equal among multiple candidates, the next dose level is chosen randomly with equal probability. The dose finding procedure for BOIN design in combination is shown in Algorithm \ref{al:BOIN-C} below. We briefly summarize the BOIN-C design here as a reference, which is a special case of the proposed BOIN-CS design in the following section. The BOIN-CS design reduces to the BOIN-C design when the selected set is equivalent to the full set of dose combinations.
\begin{algorithm}
  \caption{BOIN-C algorithm for the full set of dose combinations $\mathcal{C}_{Full}$}
  \label{al:BOIN-C}
  \begin{algorithmic}
    \State \textbf{Initialize} at dose combination $d_{11}$
    \While{the total sample size is not exhausted}
      \State Compute $\hat{\pi}_{ij}$ at the current dose $d_{\text{current}} = d_{ij}$
      \State $\mathcal{A}_e \gets \{d_{(i+1)j}, d_{i(j+1)}\} \cap \mathcal{C}_{Full}$; $\mathcal{A}_d \gets \{d_{(i-1)j}, d_{i(j-1)}\} \cap \mathcal{C}_{Full}$
      \If{$\hat{\pi}_{ij} \le \lambda_e$ \textbf{and} $\mathcal{A}_e \neq \emptyset$}
        \State $d_{\text{next}} \gets \underset{d_{i'j'} \in \mathcal{A}_e}{\arg\max}\; \hat{\pi}_{i'j'}^{\mathrm{BOIN}}$ \Comment{break ties randomly}
      \ElsIf{$\hat{\pi}_{ij} > \lambda_d$ \textbf{and} $\mathcal{A}_d \neq \emptyset$}
        \State $d_{\text{next}} \gets \underset{d_{i'j'} \in \mathcal{A}_d}{\arg\max}\; \hat{\pi}_{i'j'}^{\mathrm{BOIN}}$ \Comment{break ties randomly}
      \Else
        \State $d_{\text{next}} \gets d_{ij}$ 
      \EndIf
      \State Update $d_{\text{current}} \gets d_{\text{next}}$
    \EndWhile
  \end{algorithmic}
\end{algorithm}

During trial conduct, the BOIN framework also applies an overdosing rule: any combination $d_{ij}$ satisfying
\[
\Pr\!\bigl(\pi_{ij} > \phi \mid n_{ij}, y_{ij}\bigr) \ge \varepsilon_{\text{BOIN}}
\]
is eliminated together with all higher combinations under the assumed partial ordering. If the starting combination $d_{11}$ is eliminated, the trial stops early for safety. A more conservative safety rule may also be imposed at the starting dose in order to achieve the desired operating characteristics for the initial dose. After accrual is completed, matrix isotonic regression is applied to estimate the toxicity surface \citep{dykstra1982algorithm}, and the MTC is selected as the admissible combination whose isotonic estimate is closest to $\phi$. It is also reasonable to require that the isotonic estimate of the DLT rate at the chosen MTC is not over the de-escalation boundary. All these considerations are also applicable to the other proposed designs in this paper.

\subsection{BOIN-CS: BOIN design for selected dose combinations $\mathcal{C}_s$}
\label{sec.BOIN.CS}

In dose-escalation trials for combination therapies, it is common for at least one drug to have an established monotherapy dose level and for there to be substantial existing knowledge about the pharmacokinetic profiles, preclinical data, and pharmacodynamic data. As a result, a relatively small number of dose combinations is considered in a combination dose-escalation trial. There may be strong scientific rationale that certain combinations could yield toxicity risks that are pharmacologically implausible and exceed established safety limits, or could result in antagonistic interactions that undermine therapeutic rationale. This more targeted strategy for dose escalation can offer greater operational advantages and help speed up clinical development. Table \ref{tab:setup} illustrates several scenarios of selected dose combinations.

To accommodate the practical considerations of selected dose combinations, we generalize the BOIN combination design \citep{lin2017bayesian} to apply to any predefined subset of dose combinations, denoted as $\mathcal{C}_s \subseteq \mathcal{C}_{Full}$. This design is abbreviated as BOIN-CS. The generalized admissible sets at dose $d_{ij}$ become
\begin{align*}
\mathcal{A}_e &= \left\{d_{(i+1)j}, d_{i(j+1)}\right\} \cap \mathcal{C}_s, \\
\mathcal{A}_d &= \left\{d_{(i-1)j}, d_{i(j-1)}\right\} \cap \mathcal{C}_s,
\end{align*}
which ensures that the escalation/de-escalation only occurs among predefined acceptable dose combinations. All considerations in the BOIN framework are applied, including the optimal interval, escalation/de-escalation decisions, overdose elimination, early stopping, and final isotonic selection. 

With this extension, the BOIN-CS design exhibits the following characteristics. First, it enables the BOIN framework to be applied to a  combination dose-escalation trial with a customized dose combination space, which is usually determined based on existing monotherapy data, PK/PD information, and class effects. Second, the BOIN-CS generalizes the BOIN-C design: when $\mathcal{C}_s = \mathcal{C}_{Full}$, the BOIN-CS reduces to the BOIN-C; when $\mathcal{C}_s$ forms a single monotone path as in Table \ref{tab:setup:a} and \ref{tab:setup:b}, the BOIN-CS reduces to the BOIN design as for a monotherapy setting. Third, the BOIN-CS preserves the simplicity in implementation as a model-assisted design. Furthermore, it also preserves the theoretical property of convergence to the target dose combination as the sample size increases, with the proof shown in the Supplementary Materials. The dose finding procedure for BOIN-CS design is shown in Algorithm \ref{al:BOIN-CS}.

\begin{algorithm}
  \caption{BOIN-CS algorithm for a selected admissible subset $\mathcal{C}_s$}
  \label{al:BOIN-CS}
  \begin{algorithmic}
    \State \textbf{Initialize} at dose combination $d_{11}$
    \While{the total sample size is not exhausted}
      \State Compute $\hat{\pi}_{ij}$ at the current dose $d_{\text{current}} = d_{ij}$
      \State $\mathcal{A}_e \gets \{d_{(i+1)j}, d_{i(j+1)}\} \cap \mathcal{C}_s$; $\mathcal{A}_d \gets \{d_{(i-1)j}, d_{i(j-1)}\} \cap \mathcal{C}_s$
      \If{$\hat{\pi}_{ij} \le \lambda_e$ \textbf{and} $\mathcal{A}_e \neq \emptyset$}
        \State $d_{\text{next}} \gets \underset{d_{i'j'} \in \mathcal{A}_e}{\arg\max}\; \hat{\pi}_{i'j'}^{\mathrm{BOIN}}$ \Comment{break ties randomly}
      \ElsIf{$\hat{\pi}_{ij} > \lambda_d$ \textbf{and} $\mathcal{A}_d \neq \emptyset$}
        \State $d_{\text{next}} \gets \underset{d_{i'j'} \in \mathcal{A}_d}{\arg\max}\; \hat{\pi}_{i'j'}^{\mathrm{BOIN}}$ \Comment{break ties randomly}
      \Else
        \State $d_{\text{next}} \gets d_{ij}$
      \EndIf
      \State Update $d_{\text{current}} \gets d_{\text{next}}$
    \EndWhile
  \end{algorithmic}
\end{algorithm}

\subsection{BOIN-CE: Exploring not administered off-diagonal dose combination during de-escalation}
\label{sec.BOIN.CE}

The BOIN-CS design assumes that the preselected subset $\mathcal{C}_s$ contains all dose combinations that are clinically relevant targets. In situations where there is substantial uncertainty about which of two off-diagonal dose combinations is more toxic, we may, during dose de-escalation, prefer to assign patients to off-diagonal dose combinations that have not yet been tried. This strategy can increase the likelihood of treating patients at a wider range of dose combinations. The BOIN-CE design is introduced to address this objective; see Algorithm \ref{al:BOIN-CE} for the dose finding procedure.

Using Table~\ref{tab:setup}(d) as an illustration, suppose the current combination is A: 75 mg and B: 240 mg, and the BOIN algorithm recommends de-escalation. If a cohort of patients has been treated with the dose combination of A: 75 mg and B: 200 mg, and no patients have been treated with the other off-diagonal dose combination of A: 50 mg and B: 240 mg yet. Then, the BOIN-CE design algorithm prioritizes the unexplored dose combination of A: 50 mg and B: 240 mg. If patients have been treated with both off-diagonal dose combinations, then follow the same BOIN-CS rule.

This design affects only a narrow subset of de-escalation decisions. It can reduce the risk of missing an acceptable off-diagonal target at the cost of a slightly larger sample size.

\begin{algorithm}
  \caption{BOIN-CE algorithm with exploratory de-escalation}
  \label{al:BOIN-CE}
  \begin{algorithmic}
    \State \textbf{Initialize} at dose combination $d_{11}$
    \While{the total sample size is not exhausted}
      \State Compute $\hat{\pi}_{ij}$ at the current dose $d_{\text{current}} = d_{ij}$
      \State $\mathcal{A}_e \gets \{d_{(i+1)j}, d_{i(j+1)}\} \cap \mathcal{C}_s$; $\mathcal{A}_d \gets \{d_{(i-1)j}, d_{i(j-1)}\} \cap \mathcal{C}_s$
      \If{$\hat{\pi}_{ij} \le \lambda_e$ \textbf{and} $\mathcal{A}_e \neq \emptyset$}
        \State $d_{\text{next}} \gets \underset{d_{i'j'} \in \mathcal{A}_e}{\arg\max}\; \hat{\pi}_{i'j'}^{\mathrm{BOIN}}$ \Comment{break ties randomly}
      \ElsIf{$\hat{\pi}_{ij} > \lambda_d$ \textbf{and} $\mathcal{A}_d \neq \emptyset$}
        \State $\mathcal{U} \gets \{ d \in \mathcal{A}_d: \text{no patients treated at dose $d$} \}$ \Comment{unexplored combinations}
        \State $\mathcal{K} \gets \{ d \in \mathcal{A}_d : \text{patients treated at dose $d$} \}$ \Comment{explored combinations}
        \If{$|\mathcal{U}| = 1$ \textbf{and} $|\mathcal{K}| = 1$}
          \State $d_{\text{next}} \gets$ the unique dose combination in $\mathcal{U}$
        \Else
          \State $d_{\text{next}} \gets \underset{d_{i'j'} \in \mathcal{A}_d}{\arg\max}\; \hat{\pi}_{i'j'}^{\mathrm{BOIN}}$ \Comment{break ties randomly}
        \EndIf
      \Else
        \State $d_{\text{next}} \gets d_{ij}$ 
      \EndIf
      \State Update $d_{\text{current}} \gets d_{\text{next}}$
    \EndWhile
  \end{algorithmic}
\end{algorithm}

\subsection{BOIN-CB: BLRM-guided tie-breaking}
\label{sec.BOIN.CB}
When there is limited knowledge about the anticipated shape of the dose–toxicity relationship, using random tie-breaking is appropriate. However, if prior evidence from monotherapy studies suggests that the combination toxicity surface is smooth and can be reasonably described by a logistic function, a BLRM model may be applied to the accumulating data in the current trial to guide tie-breaking decisions instead of relying purely on randomness.

BOIN-CB is designed to retain BOIN-CS as the primary dose-assignment rule while using the BLRM only as a secondary tie-breaking device. That is, BOIN-CB invokes the BLRM only when there are ties in the posterior probabilities of the candidate dose combinations from the BOIN-CS algorithm. Let $\mathcal{D} = \{(n_{ij}, y_{ij}) : d_{ij} \in \mathcal{C}_s\}$ denote the accumulated trial data. BOIN-CB fits the BLRM to $\mathcal{D}$ and, among the BOIN-CS candidates, assigns the next cohort to the combination whose posterior mean toxicity probability under the BLRM is closest to the target $\phi$. The BLRM model specification and prior options are provided in the Supplementary Materials. When historical monotherapy data are available, \cite{schmidli2014robust} proposed a robust mixture informative prior. Although the BOIN-CB design only uses the BLRM model for handling ties, it still has the modeling assumption and requires model fitting at each dose cohort, so it is not a model-assisted design in nature. In the next section, we perform extensive simulations to evaluate the operating characteristics of the proposed designs. For each candidate dose combination $d_{i'j'}$, let $\hat{\pi}_{i'j'}^{\mathrm{BLRM}}$ denote the postrior mean DLT probability under the BLRM. The dose finding procedure for BOIN-CS design is shown in Algorithm \ref{al:BOIN-CB}.

\begin{algorithm}
  \caption{BOIN-CB algorithm with BLRM-guided tie-breaking}
  \label{al:BOIN-CB}
  \begin{algorithmic}[1]
    \State \textbf{Initialize} at dose combination $d_{11}$
    \While{the total sample size is not exhausted}
      \State Compute $\hat{\pi}_{ij}$ at the current dose $d_{\text{current}} = d_{ij}$
      \State $\mathcal{A}_e \gets \{d_{(i+1)j}, d_{i(j+1)}\} \cap \mathcal{C}_s$; $\mathcal{A}_d \gets \{d_{(i-1)j}, d_{i(j-1)}\} \cap \mathcal{C}_s$
      \If{$\hat{\pi}_{ij} \le \lambda_e$ \textbf{and} $\mathcal{A}_e \neq \emptyset$}
        \State $d_{\text{next}} \gets \underset{d_{i'j'} \in \mathcal{A}_e}{\arg\max}\; \hat{\pi}_{i'j'}^{\mathrm{BOIN}}$
        \If{$\exists\, d_{i'j'} \in \mathcal{A}_e$ with $n_{i'j'} = 0$ \textbf{or} the BOIN probabilities are tied within $\mathcal{A}_e$}
          \State $d_{\text{next}} \gets \underset{d_{i'j'} \in \mathcal{A}_e}{\arg\min}\; |\hat{\pi}_{i'j'}^{\mathrm{BLRM}} - \phi|$
        \EndIf
      \ElsIf{$\hat{\pi}_{ij} > \lambda_d$ \textbf{and} $\mathcal{A}_d \neq \emptyset$}
        \State $d_{\text{next}} \gets \underset{d_{i'j'} \in \mathcal{A}_d}{\arg\max}\; \hat{\pi}_{i'j'}^{\mathrm{BOIN}}$
        \If{$\exists\, d_{i'j'} \in \mathcal{A}_d$ with $n_{i'j'} = 0$ \textbf{or} the BOIN probabilities are tied within $\mathcal{A}_d$}
          \State $d_{\text{next}} \gets \underset{d_{i'j'} \in \mathcal{A}_d}{\arg\min}\; |\hat{\pi}_{i'j'}^{\mathrm{BLRM}} - \phi|$
        \EndIf
      \Else
        \State $d_{\text{next}} \gets d_{ij}$
      \EndIf
      \State Update $d_{\text{current}} \gets d_{\text{next}}$
    \EndWhile
  \end{algorithmic}
\end{algorithm}

\section{Simulation}
\label{sec.simulation}

A simulation study is conducted to evaluate the performance of the three proposed designs (BOIN-CS, BOIN-CB, and BOIN-CE) and to examine whether limiting the search to a prospectively justified subset can enhance both the exact and clinically acceptable MTC selection rates, as well as compare their performance in overly toxic selections.

For comparison purposes, we consider two configurations: Full dose combinations $\mathcal{C}_{Full}$ and selected subset of dose combinations $\mathcal{C}_s$.
\begin{enumerate}
\item[$\mathcal{C}_{Full}$:]
Configuration Table ~\ref{tab:setup}(c) includes the complete set of dose combinations for the two agents and represents the benchmark setting commonly used in the literature on dual-agent phase I designs.
\item[$\mathcal{C}_s$:] Configuration Table ~\ref{tab:setup}(d) includes only eight prespecified combinations forming a diagonal band. The configuration is mostly monotone but retains one pair of off-diagonal options, $(75\text{ mg}, 200\text{ mg})$ and $(50\text{ mg}, 240\text{ mg})$, so the local decision problem remains genuinely two-dimensional even after subsetting.
\end{enumerate}

In this simulation study, the target DLT probability is specified as $\phi = 0.30$. Note that alternative target DLT probabilities may be selected depending on the particular study design, such as 0.20 or 0.25. Following the standard BOIN framework, the corresponding acceptable toxicity interval is $[0.16, 0.33]$ and a dose with the true toxicity greater than 0.33 is considered as overly toxic. Each simulated trial enrolls up to 15 cohorts of size 3 and stops early if 9 patients have been treated at the current dose combination and its observed toxicity remained within the BOIN stay interval. Following the standard BOIN construction, we set $\phi_1 = 0.6\phi$ and $\phi_2 = 1.4\phi$, which yielded $(\lambda_e, \lambda_d) = (0.236, 0.359)$. The overdosing cutoff was set to $\epsilon_{\mathrm{BOIN}} = 0.95$.

 Table~\ref{tab:toxicity-scenarios} shows the 14 toxicity scenarios that are  constructed analogously to those in \cite{barnett2024comparison}. They can be grouped into four clinically interpretable families. Scenarios 1--4 represent relatively regular surfaces with a single MTC on or near the upper boundary of the clinically relevant region. Scenarios 5--7 place the MTC in the interior or along a short target ridge, so the design must distinguish among several nearby admissible combinations. Scenario 8 is an irregular setting in which toxicity jumps as the dose level of drug A increases from 3 to 4. Scenarios 9--13 are predominantly toxic settings in which only a small lower-left region remains clinically acceptable. Scenario 14 is the ethical null case in which every combination is overly toxic and the correct practical action is to recommend no dose combination.

For each scenario, we generated 1000 Monte Carlo trials. We summarize performance using five operating characteristics: the proportion of correct selections (PCS), the proportion of acceptable selections (PAS), the proportion of overly toxic selections, the average number of patients treated in total, and the average number of DLTs. PCS measures the identification of the true MTC, PAS measures whether the final recommendation remains clinically usable even when not exactly at the target, the proportion of overly toxic selections measures recommendation safety, and the patient-allocation and DLT summaries characterize patient safety and trial burden. The first three operating characteristics are summarized in this section. The other two are included in the Supplementary Materials. In addition, the simulation details for the configuration of the full set of dose combinations $\mathcal{C}_{Full}$ are included in the Supplementary Materials. For BOIN-CB, vague priors were used for all scenarios.

\begin{table}[htbp]
\centering
\begin{tabular}{c|cccc|cccc}
\toprule
\textbf{Drug A}
  & \multicolumn{8}{c}{\textbf{Drug B}}
 \\
 & $d^B_1$ & $d^B_2$ & $d^B_3$ & $d^B_4$
   & $d^B_1$ & $d^B_2$ & $d^B_3$ & $d^B_4$ \\
\midrule
 & \multicolumn{4}{c|}{\textbf{Scenario 1}} & \multicolumn{4}{c}{\textbf{Scenario 2}} \\
$d^A_1$ & 0.05 & 0.08 & 0.12 & 0.15 & 0.05 & 0.07 & 0.10 & 0.15 \\
$d^A_2$ & 0.08 & 0.12 & 0.15 & 0.20 & 0.07 & 0.10 & 0.15 & 0.20 \\
$d^A_3$ & 0.12 & 0.15 & 0.23 & 0.25 & 0.10 & 0.15 & 0.20 & \textbf{0.30} \\
$d^A_4$ & 0.15 & 0.20 & 0.25 & \textbf{0.30} & 0.15 & 0.20 & \textbf{0.30} & 0.45 \\
\midrule
 & \multicolumn{4}{c|}{\textbf{Scenario 3}} & \multicolumn{4}{c}{\textbf{Scenario 4}} \\
$d^A_1$ & 0.02 & 0.05 & 0.08 & 0.10 & 0.05 & 0.08 & 0.12 & 0.15 \\
$d^A_2$ & 0.05 & 0.12 & 0.15 & 0.20 & 0.10 & 0.15 & 0.20 & 0.25 \\
$d^A_3$ & 0.10 & 0.15 & 0.20 & 0.25 & 0.15 & 0.20 & 0.25 & \textbf{0.30} \\
$d^A_4$ & 0.20 & 0.25 & \textbf{0.30} & 0.45 & 0.20 & 0.40 & 0.50 & 0.60 \\
\midrule
 & \multicolumn{4}{c|}{\textbf{Scenario 5}} & \multicolumn{4}{c}{\textbf{Scenario 6}} \\
$d^A_1$ & 0.02 & 0.04 & 0.08 & 0.15 & 0.10 & 0.15 & \textbf{0.30} & 0.45 \\
$d^A_2$ & 0.07 & 0.10 & 0.15 & 0.25 & 0.15 & \textbf{0.30} & 0.45 & 0.55 \\
$d^A_3$ & 0.15 & 0.20 & \textbf{0.30} & 0.63 & \textbf{0.30} & 0.45 & 0.55 & 0.58 \\
$d^A_4$ & 0.40 & 0.45 & 0.63 & 0.65 & 0.45 & 0.55 & 0.58 & 0.60 \\
\midrule
 & \multicolumn{4}{c|}{\textbf{Scenario 7}} & \multicolumn{4}{c}{\textbf{Scenario 8}} \\
$d^A_1$ & 0.10 & 0.15 & 0.25 & 0.45 & 0.05 & 0.10 & 0.15 & 0.20 \\
$d^A_2$ & 0.12 & 0.20 & 0.45 & 0.50 & 0.08 & 0.12 & 0.20 & 0.25 \\
$d^A_3$ & 0.15 & \textbf{0.30} & 0.50 & 0.55 & 0.13 & 0.14 & 0.22 & \textbf{0.30} \\
  $d^A_4$ & \textbf{0.30} & 0.40 & 0.55 & 0.60 & \textbf{0.30} & 0.40 & 0.50 & 0.55 \\
\midrule
 & \multicolumn{4}{c|}{\textbf{Scenario 9}} & \multicolumn{4}{c}{\textbf{Scenario 10}} \\
$d^A_1$ & 0.10 & 0.22 & \textbf{0.30} & 0.45 & 0.15 & \textbf{0.30} & 0.45 & 0.55 \\
$d^A_2$ & 0.22 & 0.35 & 0.50 & 0.58 & \textbf{0.30} & 0.45 & 0.55 & 0.60 \\
$d^A_3$ & \textbf{0.30} & 0.50 & 0.58 & 0.60 & 0.45 & 0.55 & 0.60 & 0.63 \\
$d^A_4$ & 0.45 & 0.58 & 0.60 & 0.60 & 0.55 & 0.60 & 0.63 & 0.65 \\
\midrule
 & \multicolumn{4}{c|}{\textbf{Scenario 11}} & \multicolumn{4}{c}{\textbf{Scenario 12}} \\
$d^A_1$ & 0.02 & 0.05 & 0.10 & 0.15 & 0.20 & \textbf{0.30} & 0.45 & 0.60 \\
$d^A_2$ & 0.08 & 0.25 & 0.45 & 0.60 & 0.35 & 0.50 & 0.60 & 0.70 \\
$d^A_3$ & \textbf{0.30} & 0.40 & 0.50 & 0.70 & 0.45 & 0.60 & 0.70 & 0.73 \\
$d^A_4$ & 0.45 & 0.60 & 0.70 & 0.75 & 0.60 & 0.70 & 0.73 & 0.75 \\
\midrule
 & \multicolumn{4}{c|}{\textbf{Scenario 13}} & \multicolumn{4}{c}{\textbf{Scenario 14}} \\
$d^A_1$ & \textbf{0.30} & 0.40 & 0.50 & 0.55 & 0.45 & 0.50 & 0.55 & 0.60 \\
$d^A_2$ & 0.40 & 0.50 & 0.55 & 0.58 & 0.50 & 0.55 & 0.60 & 0.63 \\
$d^A_3$ & 0.50 & 0.55 & 0.58 & 0.60 & 0.55 & 0.60 & 0.63 & 0.65 \\
$d^A_4$ & 0.55 & 0.58 & 0.60 & 0.60 & 0.60 & 0.63 & 0.65 & 0.65 \\
\bottomrule
\end{tabular}
\caption{True toxicity probabilities for the full $4\times4$ dose matrix under 14 scenarios. Bold entries identify the MTC(s).}
\label{tab:toxicity-scenarios}
\end{table}

\subsection*{Proportion of correct and acceptable selections}

\begin{figure}[htbp]
    \centering
    \includegraphics[width=0.90\linewidth]{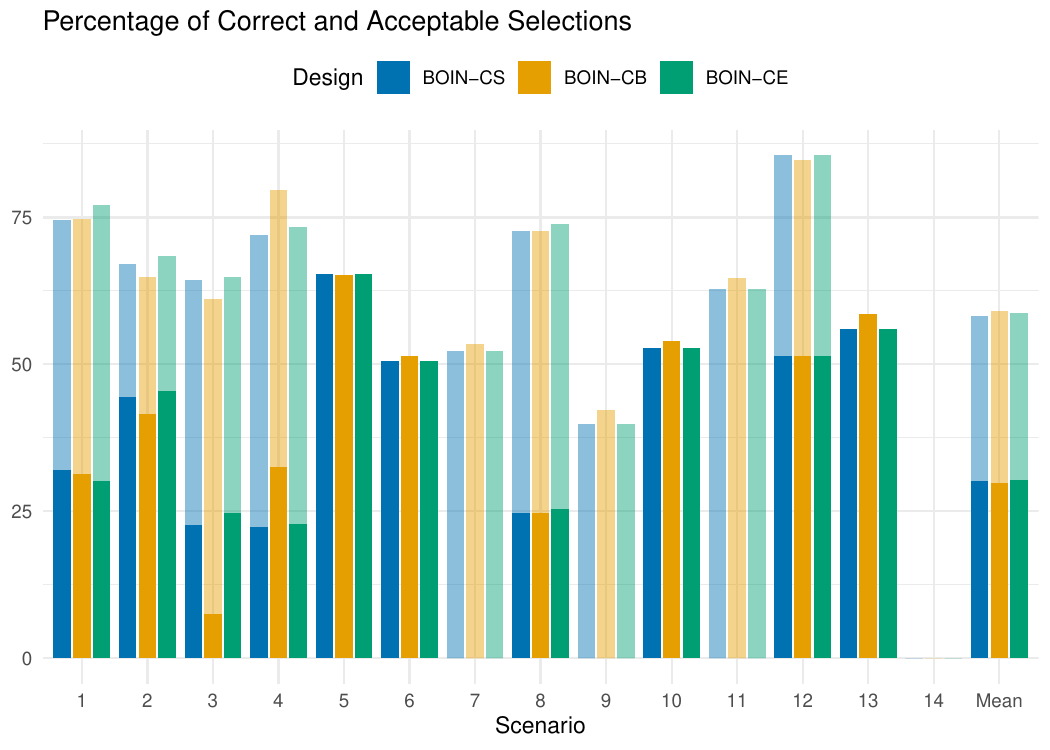}
    \caption{Proportion of correct selections (PCS) and proportion of acceptable selections (PAS) for Scenarios 1--14 for the selected configuration $\mathcal{C}_s$. Solid bars show PCS, translucent bars show PAS, and the cluster at the far right summarizes the overall average across all 14 scenarios.}
    \label{fig:MTD_select}
\end{figure}

Figure~\ref{fig:MTD_select} displays the proportion of correct selections (PCS) and the proportion of acceptable selections (PAS). The three designs have similar average PCS across the 14 scenarios: 30.17\%, 29.88\%, and 30.34\% for BOIN-CS, BOIN-CB, and BOIN-CE, respectively. In addition, the average acceptable selection rates are also similar: 58.29\%, 59.09\%, and 58.78\% for the three designs, respectively.

When the prespecified subset of dose combinations is well aligned with the clinically relevant target region (e.g., Scenario 5), the restricted search improves both accuracy and safety. In this scenario, BOIN-CS achieved a PCS of 65.4\% and an overly toxic recommendation rate of 5.2\%. Scenario 4 illustrates when BLRM-guided tie-breaking can help. BOIN-CB increased PCS from 22.3\% under BOIN-CS and 22.8\% under BOIN-CE to 32.5\%, increased PAS to 79.6\%, and reduced overly toxic selection to 4.1\%. Scenario 8 shows the more limited but still visible benefit of exploratory recovery: BOIN-CE increased PAS to 73.9\% versus 72.7\% for BOIN-CS and BOIN-CB, while slightly lowering overly toxic selection from 13.7\% to 12.5\%. On the other hand, if the target dose combination is excluded from the prespecified subset of dose combinations (e.g., Scenarios 9 and 14), the PCS is 0\%. 

% \begin{table}[htbp]
% \centering
% \caption{Prior mean toxicity probabilities for the BLRM across dose combinations of drugs A and B.}
% \label{tab:simu_blrm_tox}
% \begin{tabular}{rcccc}
% \toprule
%  & \multicolumn{4}{c}{\textbf{Drug B (mg)}}\\
% \cmidrule(lr){2-5}
% \textbf{Drug A (mg)} & \textbf{120} & \textbf{160} & \textbf{200} & \textbf{240} \\
% \midrule
% \textbf{15}  & 0.27 & 0.32 & 0.36 & 0.39 \\
% \textbf{25}  & 0.31 & 0.36 & 0.39 & 0.43 \\
% \textbf{50} & 0.40 & 0.44 & 0.47 & 0.50 \\
% \textbf{75} & 0.47 & 0.50 & 0.53 & 0.56\\
% \bottomrule
% \end{tabular}
% \end{table}

\subsection*{Effect of restricting the search space}

\begin{table}[htbp]
\centering
\begin{tabular}{lccc}
\toprule
Method & PCS (\%) & PAS (\%) & \shortstack[c]{Overly toxic\\selection (\%)} \\
\midrule
BOIN-C on $\mathcal{C}_{\mathrm{full}}$ & 30.94 & 55.83 & 27.35 \\
BOIN-CS on $\mathcal{C}_s$ & 30.17 & 58.29 & 21.70 \\
\bottomrule
\end{tabular}
\caption{Mean operating characteristics across 14 scenarios for BOIN-C applied to the full dose-combination set, $\mathcal{C}_{Full}$, and BOIN-CS applied to the prespecified subset, $\mathcal{C}_s$.}
\label{tab:subset_benchmark}
\end{table}

Table~\ref{tab:subset_benchmark} compares BOIN-C on PCS, PAS, and the overly toxic selection rate for the full set of dose combinations $\mathcal{C}_{Full}$ versus BOIN-CS for the subset of dose combinations $\mathcal{C}_s$. Because the two designs operate on different admissible search spaces, this comparison is intended to quantify the practical effect of prospectively restricting exploration rather than to claim algorithmic superiority. On average, restricting the search space preserved PCS, improved clinically acceptable selection, and reduced the overly toxic recommendation rate.

When the selected subset contained the clinically relevant region while excluding implausible higher-toxicity combinations, the benefit of BOIN-CS is apparent. For instance, in Scenario 5, BOIN-CS more than doubled the PCS compared to BOIN-C (65.4\% vs. 29.2\%) while simultaneously lowering the rate of overly toxic selections (5.2\% vs. 19.0\%). Similarly, in Scenario 12, it increased PAS (85.6\% vs. 53.7\%) and enhanced recommendation safety (11.0\% vs. 43.6\%).

\subsection*{Proportion of overly toxic selections}

\begin{figure}[htbp]
    \centering
    \includegraphics[width=0.90\linewidth]{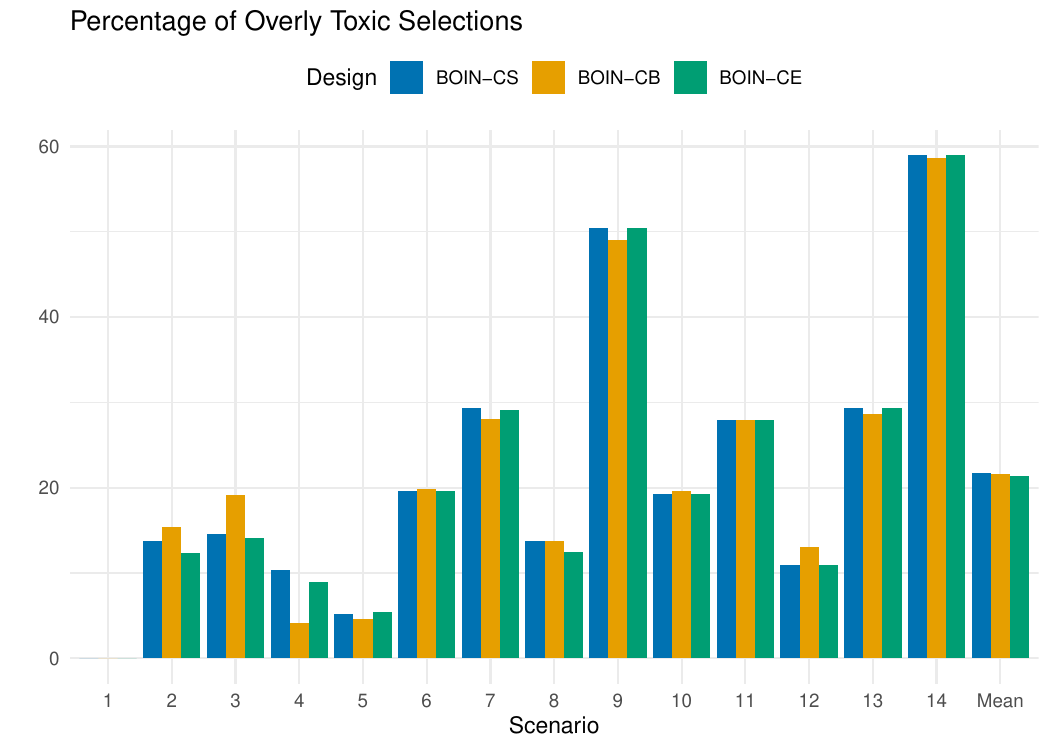}
    \caption{Proportion of overly toxic selections for Scenarios 1--14 for the selected configuration $\mathcal{C}_s$. The cluster at the far right summarizes the overall average across all 14 scenarios}
    \label{fig:overlytoxicselection}
\end{figure}

Figure~\ref{fig:overlytoxicselection} displays the proportion of overly toxic selections for configuration $\mathcal{C}_s$. The averaged overly toxic selection rate across all 14 scenarios is very similar: 21.70\% , 21.56\% , and 21.37\% for BOIN-CS, BOIN-CB, and BOIN-CE, respectively.

When the prespecified subset of dose combinations is well aligned with the clinically relevant target region (e.g., Scenario 5), BOIN-CS has a low  overly toxic recommendation rate of 5.2\%. Scenario 4 shows that BOIN-CB can further improve safety when the BLRM model favors the correct region, reducing the overly toxic selection rate to 4.1\%. Scenario 8 shows a smaller exploratory benefit, with BOIN-CE reducing the overly toxic recommendation rate from 13.7\% to 12.5\%. 

Additional summaries, including the number of patients treated at excessively toxic dose combinations, the mean number of DLTs, and the achieved sample size, are provided in the Supplementary Materials.

\section{A case study}
\label{sec.case}

Consider a hypothetical phase I dual-agent trial where drug A has four dose levels (15 mg, 25 mg, 50 mg, and 75 mg) and drug B also has four dose levels (120 mg, 160 mg, 200 mg, and 240 mg). The complete set of dose combinations consists of 16 in total, but only 8 combinations in Table~\ref{tab:case_subset} ($\mathcal{C}_{s}$) are considered in the dose escalation study based on prior monotherapy safety and pharmacology data. Suppose the true toxicity profile is Scenario 8 from Table~\ref{tab:toxicity-scenarios}, which contains a single MTC included in $\mathcal{C}_{s}$.

The true toxicity probabilities for the selected subset are shown in Table~\ref{tab:case_subset}. The target combination is $(50\text{ mg}, 240\text{ mg})$, whose toxicity probability is 0.30. In $\mathcal{C}_{s}$, the dose combination $(75\text{ mg}, 200\text{ mg})$ is excluded a priori because it is regarded as clinically undesirable. The admissible search path, therefore, retains a local branching decision at the intermediate doses, while the final approach to the target region proceeds through a predetermined escalation path.

%\begin{table}[htbp]
%\centering
%\caption{Prespecified admissible dose spaces $\mathcal{C}_{s}$ for the case study. Gray cells indicate combinations included in the trial.}
%\label{tab:setup_case}
%\newcommand{\GridHeader}{%
%  & \multicolumn{4}{c}{\textbf{Drug B (mg)}} \\ \cmidrule(lr){2-5}
%  \textbf{Drug A (mg)} & 120 & 160 & 200 & 240 \\ \midrule
%}
%  \centering
%  \begin{tabular}{rcccc}
%    \toprule
%    \GridHeader
%     15 & \on  & \on  & \off & \off \\
%     25 & \off & \on  & \on  & \off \\
%     50 & \off & \on & \on  & \on  \\
%     75 & \off & \off & \off  & \on  \\ \bottomrule
%  \end{tabular}
%\end{table}

\begin{table}[htbp]
\centering
\begin{tabular}{rcccc}
\toprule
 & \multicolumn{4}{c}{\textbf{Drug B (mg)}}\\
\cmidrule(lr){2-5}
\textbf{Drug A (mg)} & \textbf{120} & \textbf{160} & \textbf{200} & \textbf{240} \\
\midrule
\textbf{15} & 0.05 & 0.10 & -- & -- \\
\textbf{25} & -- & 0.12 & 0.20 & -- \\
\textbf{50} & -- & 0.14 & 0.22 & \textbf{0.30} \\
\textbf{75} & -- & -- & -- & 0.55 \\
\bottomrule
\end{tabular}
\caption{True toxicity probabilities for the selected subset $\mathcal{C}_{s}$ used in the case study. Dashes indicate combinations excluded from the trial by design. The true MTC is shown in bold (Drug A: 50 mg, Drug B: 240 mg)}
\label{tab:case_subset}
\end{table}

\begin{figure}
    \centering
    \includegraphics[width=0.75\linewidth]{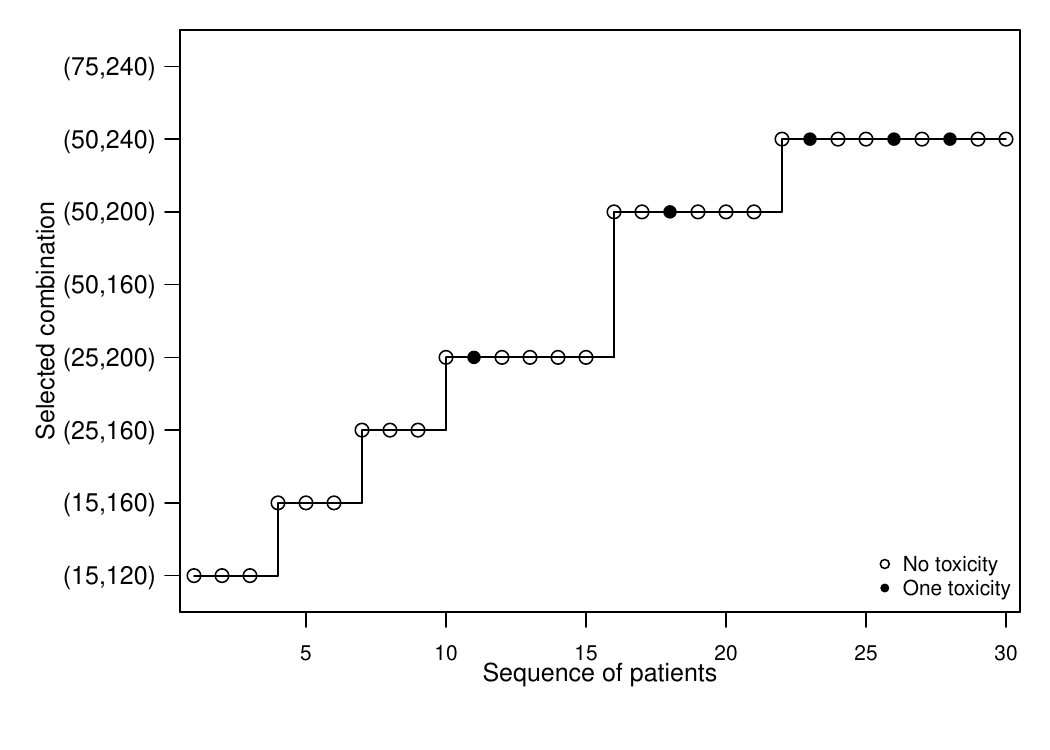}
    \caption{Illustrative BOIN-CS trial trajectory under Scenario~8. Open circles denote patients without DLTs and solid circles denote patients with DLTs.}
    \label{fig:casestudy}
\end{figure}

The trial enrolls patients in cohorts of 3 and allows a maximum of 15 cohorts. As in the simulation study, we target $\phi = 0.30$ and set $(\phi_1,\phi_2)=(0.6\phi,1.4\phi)$, which yields $(\lambda_e,\lambda_d)=(0.236,\,0.359)$. The BOIN overdosing cutoff is set to $\epsilon_{\mathrm{BOIN}} = 0.90$. The first cohort is treated at the lowest admissible combination, $(15\text{ mg}, 120\text{ mg})$. Figure~\ref{fig:casestudy} provides a schematic BOIN-CS trajectory under this scenario, with the observed DLTs and the number of patients treated at each dose combination.

Since no DLTs are observed in the first 3 cohorts of patients, BOIN-CS escalates from $(15\text{ mg}, 120\text{ mg})$ to $(15\text{ mg}, 160\text{ mg})$ and then to $(25\text{ mg}, 160\text{ mg})$. The first branching decision arises at $(25\text{ mg}, 160\text{ mg})$, where the admissible escalation set $\mathcal{A}_e$ contains two combinations, $(50\text{ mg}, 160\text{ mg})$ and $(25\text{ mg}, 200\text{ mg})$. By random selection with equal probability, the design escalates to $(25\text{ mg}, 200\text{ mg})$, where 1 DLT is observed out of 6 patients. BOIN-CS continues escalating to $(50\text{ mg}, 200\text{ mg})$. Because $(75\text{ mg}, 200\text{ mg})$ is not part of the admissible subset, any additional escalation from $(50\text{ mg}, 200\text{ mg})$ can proceed only to $(50\text{ mg}, 240\text{ mg})$, the true MTC. After observing 3 DLTs among 9 treated patients at that combination, the observed toxicity rate was 0.333, which lay within the BOIN stay interval. The design therefore retained that combination and stopped early after accumulating 9 patients at the current dose, with observed toxicity within the stay interval.

The final observed toxicity rates on the explored combinations were $0/3$, $0/3$, $0/3$, $1/6$, $1/6$, and $3/9$ at $(15\text{ mg},120\text{ mg})$, $(15\text{ mg},160\text{ mg})$, $(25\text{ mg},160\text{ mg})$, $(25\text{ mg},200\text{ mg})$, $(50\text{ mg},200\text{ mg})$, and $(50\text{ mg},240\text{ mg})$, respectively. The isotonic estimate preserved the monotone pattern along the explored band and selected $(50\text{ mg}, 240\text{ mg})$ as the MTC. This example illustrates the main practical advantage of the BOIN-CS design by exploring a meaningful prespecified subset of dose combinations.

\section{Discussion}
\label{sec.discussion}
The dose escalation trials for combination therapies usually include a small number of targeted dose combinations rather than the full Cartesian dose combinations. The existing statistical designs, such as BOIN-C, cannot be directly applied. In this paper, three approaches (BOIN-CS, BOIN-CB, and BOIN-CE) are proposed as extensions to the BOIN-C design that can address a customized subset of dose combinations. Their asymptotic property is also shown to still hold. The simulation study explored a wide range of toxicity profiles and demonstrated the utility of the proposed methods. These gains are practically important: in phase I combination trials, every unnecessary exploration of an implausible region of the dose combinations consumes unnecessary resources and may expose patients to avoidable risk. The three proposed methods differ in their specific characteristics. When a tie occurs, BOIN-CS chooses a dose combination at random with equal probability, whereas BOIN-CB relies on the available data through the BLRM model to break the tie. 
When two dose combinations are available for de-escalation, BOIN-CE prefers moving to the previously unexplored combination if one exists. In contrast, BOIN-CS selects between the two by comparing their posterior probabilities, where the probability for the unexplored combination is calculated using a vague prior; and BOIN-CB determines the de-escalation dose by the BLRM model.
All three proposed extensions rely on the assumption that the chosen subset of dose combinations is clinically meaningful and that the target dose combination is either contained within this subset or does not exist anywhere in the full Cartesian set of dose combinations. If there is substantial uncertainty about whether the selected subset contains the true target region, then expanding the selected subset of dose combinations may be necessary. Ultimately, the restricted subset of dose combinations requires adequate scientific and clinical justification. In addition, it is worth noting that the BOIN-CB design incorporates a BLRM component that relies on a bivariate logistic model assumption. Although this BLRM component is used solely to resolve ties, its underlying modeling assumption still affects dose selection. Therefore, it remains important to assess whether this assumption is reasonable. The selection of the most suitable design among the three proposed approaches is determined by the particular study context, taking into account existing clinical and preclinical data, current knowledge of class-related toxicities, and the PK/PD characteristics derived from prior clinical and preclinical studies. In addition, comprehensive simulation studies are necessary to evaluate the operating characteristics of the chosen design for the specific trial.

In this paper, we focus on toxicity-guided dose escalation for combinations of two agents and do not account for efficacy, delayed toxicity or backfilling. Future research could broaden these methods to utility-based dose finding that jointly considers toxicity and efficacy, explicitly models late-onset toxicity, and incorporates backfilled patients into the dose-escalation process.

\bibliography{boin-ref}
\bibliographystyle{plainnat}

\end{document}

\begin{center}
{\Large Supplementary Materials}\\[0.4em]
\end{center}

\subsection*{S1. The BLRM model used in BOIN-CB for breaking ties}
\label{app:BLRM}
Let $\pi_{ij}$ be the DLT rate of the dose combination of Drug A dose level $i$ and Drug B dose level $j$, where $i=1,2,\cdots,I$ and $j=1,2,\cdots, J$. The dose levels are scaled relative to the highest dose level in each drug: $d_i^{A*} = d_i^A/d_I^A$ and $d_j^{B*} = d_j^B/d_J^B$. In the BOIN-CB design, when there are ties, the dual-agent BLRM model \citep{neuenschwander2015bayesian} below is used to break the ties:
\begin{equation}
    \text{logit}(\pi_{ij}) = \log\!\left[\alpha_1 (d_i^{A*})^{\beta_1} + \alpha_2 (d_j^{B*})^{\beta_2} + \alpha_1 \alpha_2 (d_i^{A*})^{\beta_1}(d_j^{B*})^{\beta_2} \right] + \eta d_i^{A*} d_j^{B*},\nonumber
\end{equation}
where $\alpha_1$ and $\beta_1$ are used to characterize the marginal toxicity of drug A; $\alpha_2$ and $\beta_2$ are used to characterize the marginal toxicity of drug B; and $\eta$ is for the interaction between drug A and drug B. The model incorporates all existing data from previous and current dose combinations through a smooth parametric toxicity surface. In the BOIN-CB design, however, the BLRM model is used only to break ties or to select among the candidate dose combinations that have not been used to treat patients yet. The other escalation and de-escalation decisions are the same as in the BOIN-CS design.

Following \cite{neuenschwander2015bayesian}, the weakly informative prior is
\begin{equation*}
    (\log(\alpha_k), \log(\beta_k)) \sim \mathrm{BVN}\!\bigl((\mu_{\alpha k}, \mu_{\beta k}), \Sigma_k\bigr), \qquad \text{for } k = A, B,
\end{equation*}
with $\mu_{\alpha k} = \mathrm{logit}(p_k^*)$, $\mu_{\beta k} = 0$, and
$
\Sigma_k = \begin{bmatrix}
2 & 0 \\
0 & 1
\end{bmatrix},
$
where $p_A^*$ and $p_B^*$ denote the anticipated DLT rates at the scaling dose levels $d_I^{A*}$ and $d_I^{B*}$ for drug $A$ and $B$, respectively. When historical monotherapy data are available, informative or robust mixture priors can be substituted to borrow information while guarding against prior-data conflict \citep{schmidli2014robust}.
In the simulation study, weakly informative priors were specified with $p_k^* = 0.33$ for each drug and $\eta \sim N(0, 1.121^2)$. The zero mean permits both synergistic and antagonistic interactions, and the variance is sufficiently large to prevent imposing a strong prior at the scaled dose levels.

\subsection*{S2. Asymptotic Property}
Assume partial monotonicity holds on the admissible set $\mathcal{C}_s$ and any two dose combinations in $\mathcal{C}_s$ are connected through adjacent admissible moves. Then BOIN-CS has the same large-sample property as BOIN-C on the partially ordered set: if at least one admissible combination has true toxicity in $(\lambda_e, \lambda_d)$, the allocation converges almost surely to one such combination; otherwise, it converges to admissible boundary combinations that bracket the target interval.

The result above is obtained by applying Theorem 2.1 in \cite{lin2017bayesian} to the restricted admissible set $\mathcal{C}_s$. BOIN-CS modifies only the collection of permissible local moves; it leaves the BOIN interval-based decision rule unchanged at every admissible combination. BOIN-CB differs from BOIN-CS solely in its finite-sample tie-breaking, and the probability of observing an exact tie goes to zero as the number of observations at the candidate combinations grows. BOIN-CE may occasionally compel exploration of an untested admissible combination, but this can happen only a finite number of times because $\mathcal{C}_s$ contains only finitely many elements. Therefore, both BOIN-CB and BOIN-CE exhibit the same asymptotic property as BOIN-CS.

\section*{S3. Simulation Results for Configuration $\mathcal{C}_{Full}$}

This section provides the scenario-by-scenario benchmark results under the full $4\times 4$ dose combinations. 
\begin{figure}[H]
    \centering
    \includegraphics[width=0.90\linewidth]{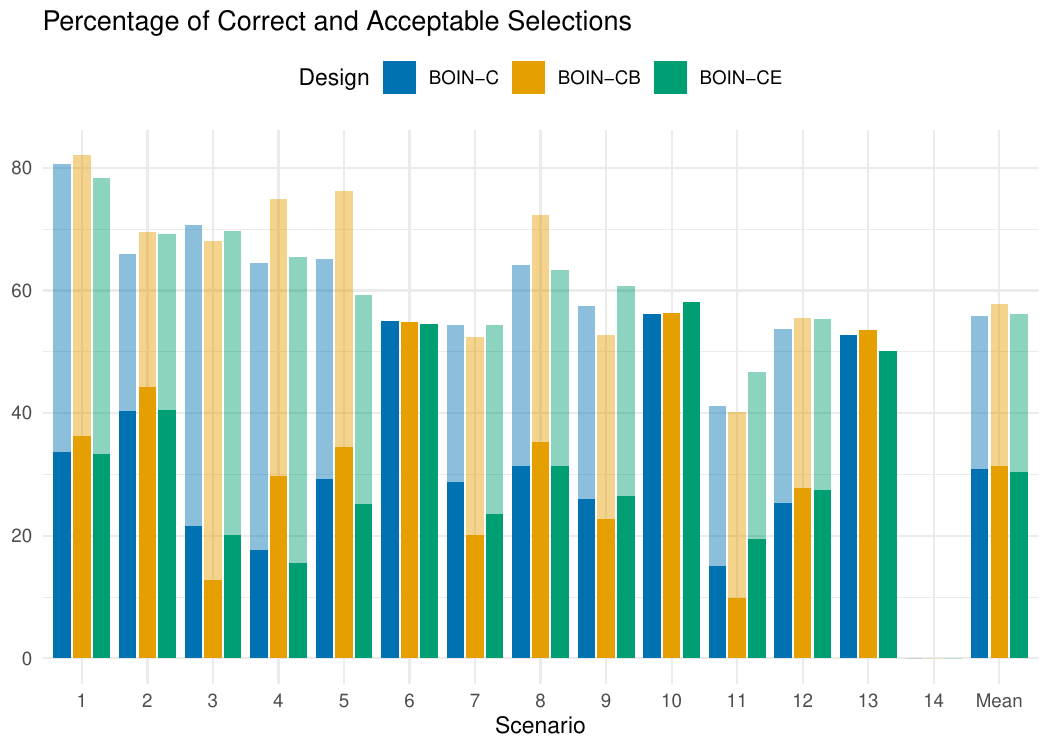}
    \caption{Proportion of correct selections (PCS) and proportion of acceptable selections (PAS) for Scenarios 1--14 under the full configuration $\mathcal{C}_{\mathrm{Full}}$. Solid bars show PCS, translucent bars show PAS, and the cluster at the far right summarizes the overall average across all 14 scenarios.}
    \label{fig:supp_full_primary}
\end{figure}

\begin{figure}[H]
    \centering
    \includegraphics[width=0.90\linewidth]{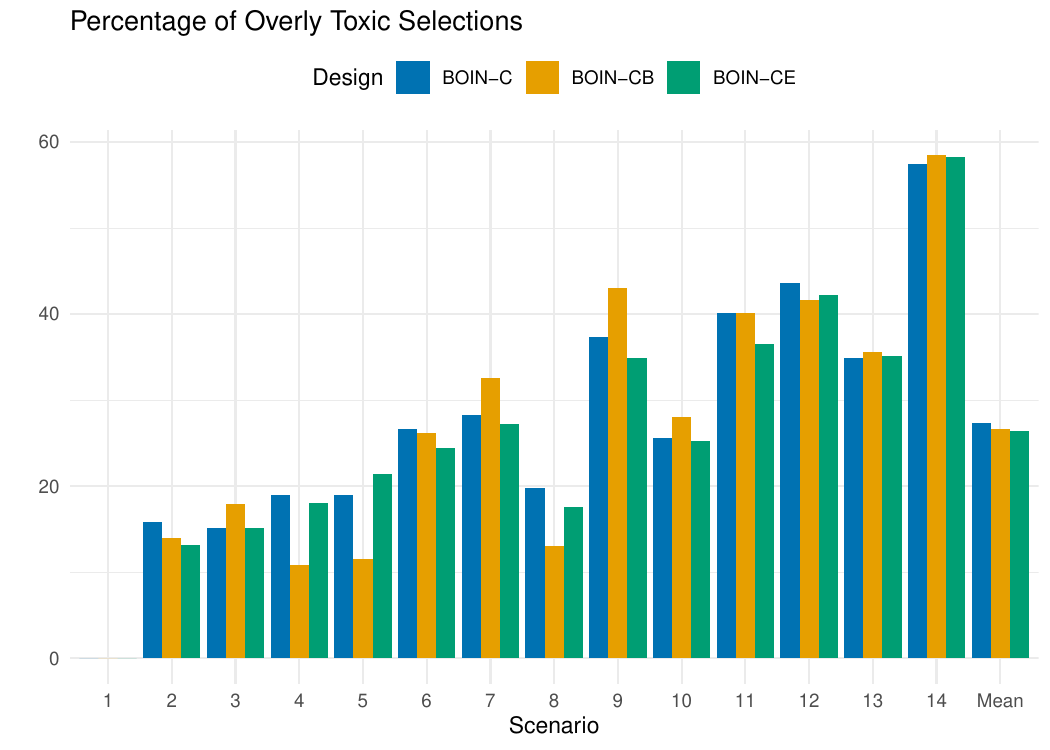}
    \caption{Proportion of overly toxic selections for Scenarios 1--14 under the full configuration $\mathcal{C}_{\mathrm{Full}}$. The cluster at the far right summarizes the overall average across all 14 scenarios.}
    \label{fig:supp_full_overdose}
\end{figure}

Under the full grid, the overall pattern was similar to the selected-grid analysis but less favorable on average when the true target lay within the prespecified selected band. Mean PAS was 55.83\% for BOIN-C on the full grid, compared with 58.29\% for BOIN-CS on the selected grid, and mean overly toxic selection was 27.35\% for BOIN-C, compared with 21.70\% for BOIN-CS.

\section*{S4. Additional Measures of Operating Characteristics}

Additional operating characteristics are summarized in Tables~\ref{tab:supp-selected-secondary} including the realized sample size, the total number of DLTs, and the average number of patients treated at excessively toxic dose combinations. 
\begin{figure}[H]
    \centering
    \begin{subfigure}[t]{0.90\linewidth}
        \centering
        \includegraphics[width=\linewidth]{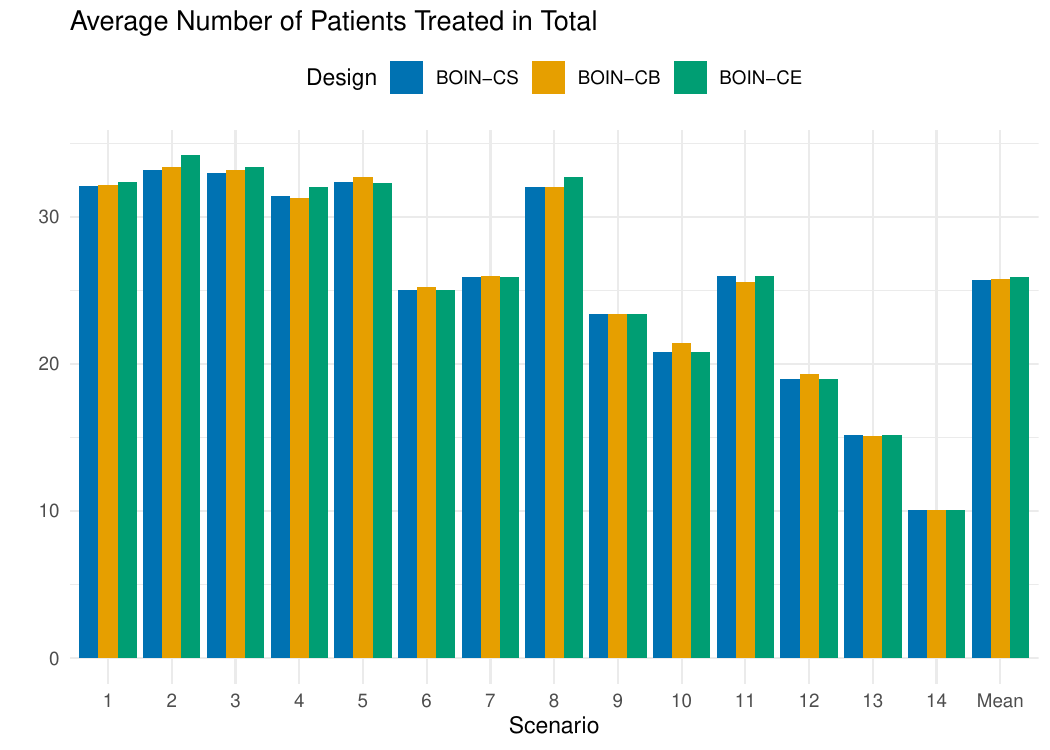}
        \caption{Selected Subset of Dose Combinations $\mathcal{C}_s$ }
    \end{subfigure}

    \vspace{0.8em}

    \begin{subfigure}[t]{0.90\linewidth}
        \centering
        \includegraphics[width=\linewidth]{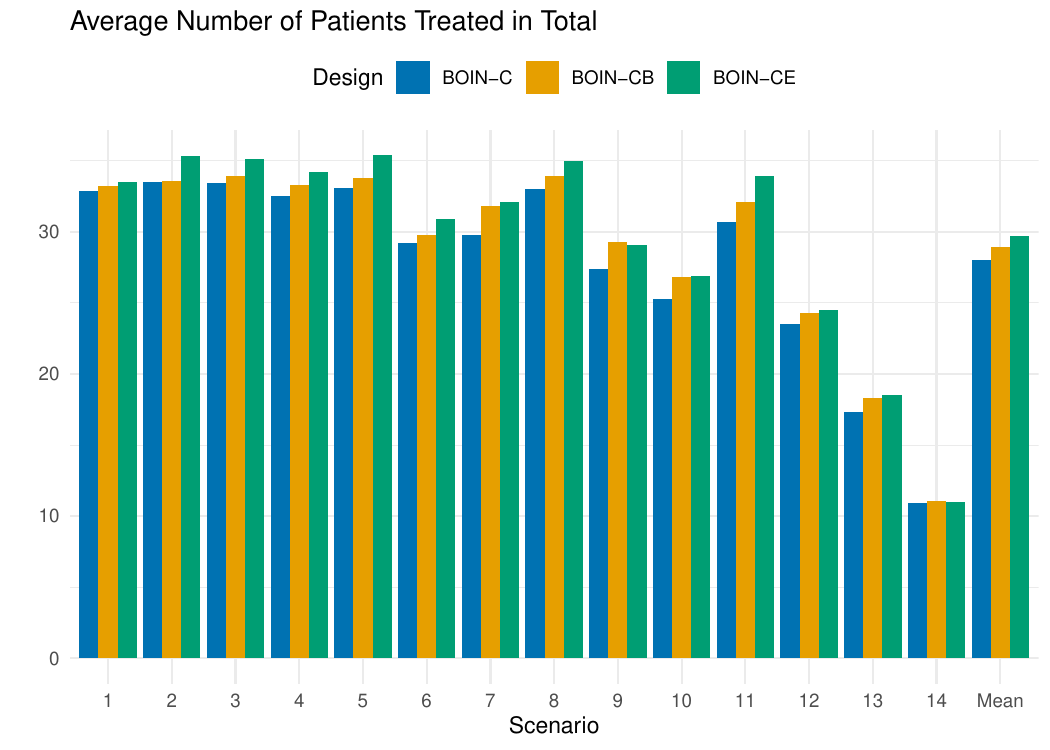}
        \caption{Full Set of Dose Combinations $\mathcal{C}_{\mathrm{Full}}$}
    \end{subfigure}
    \caption{Average number of patients treated for Scenarios 1--14. The cluster at the far right summarizes the overall average across all 14 scenarios.}
    \label{fig:supp_samplesize}
\end{figure}

\begin{figure}[H]
    \centering
    \begin{subfigure}[t]{0.90\linewidth}
        \centering
        \includegraphics[width=\linewidth]{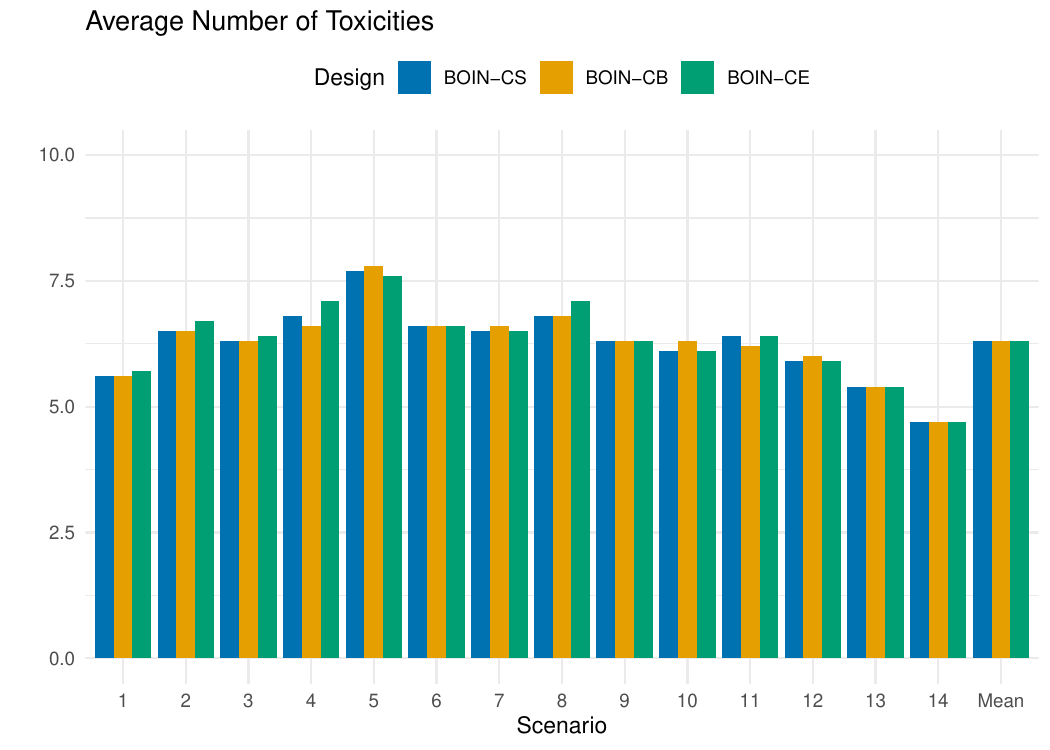}
        \caption{Selected Subset of Dose Combinations $\mathcal{C}_s$ }
    \end{subfigure}

    \vspace{0.8em}

    \begin{subfigure}[t]{0.90\linewidth}
        \centering
        \includegraphics[width=\linewidth]{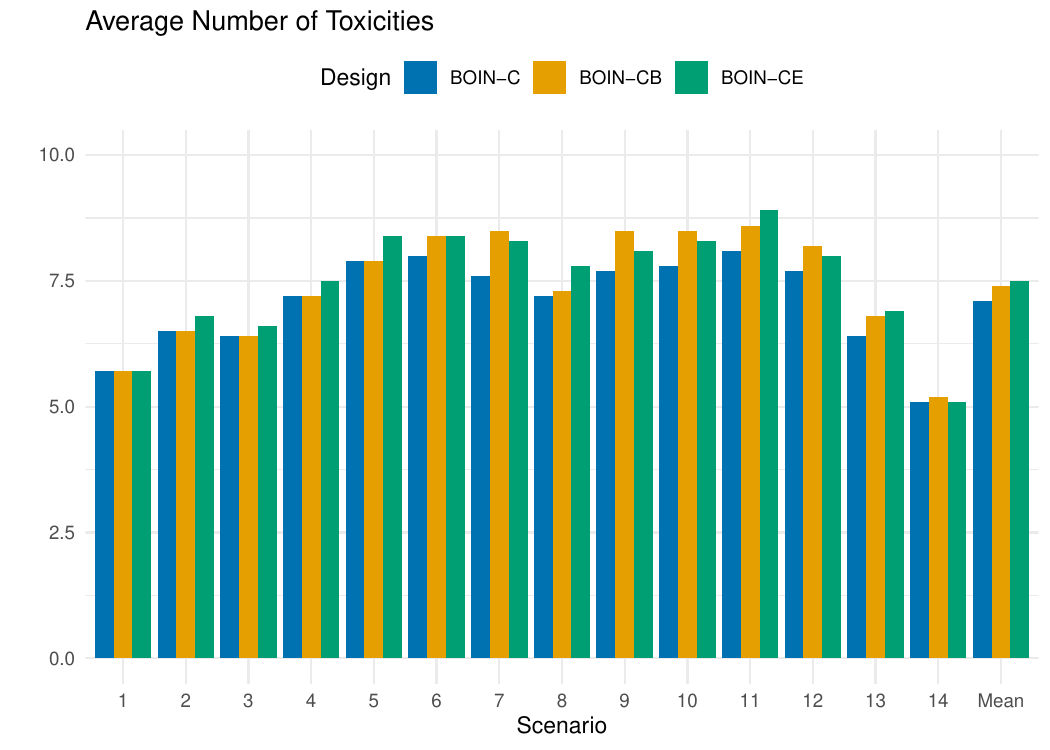}
        \caption{Full Set of Dose Combinations $\mathcal{C}_{\mathrm{Full}}$}
    \end{subfigure}
    \caption{Average number of DLTs for Scenarios 1--4. The cluster at the far right summarizes the overall average across all 14 scenarios.}
    \label{fig:supp_toxicities}
\end{figure}

The configuration of selected subset of dose combinations $\mathcal{C}_s$, as expected, reduced the average sample size and average total DLTs compared to those using the full set of dose combinations $\mathcal{C}_{\mathrm{Full}}$. Across all 14 scenarios in the simulation study, the average realized sample size decreased from 28.9 under $\mathcal{C}_{\mathrm{Full}}$ to 25.8 under $\mathcal{C}_s$, while the mean number of DLTs decreased from 7.3 to 6.3. These gains in conduct safety and efficiency were most evident in the predominantly toxic scenarios, where restricting the admissible set prevented repeated exploration of clearly implausible regions of the dose combinations.

\section*{Summaries of the Operating Characteristics}

\begin{table}[H]
\centering
\caption{Operating Characteristics: PCS, PAS, and Proportion of Overly Toxic Selection}
\label{tab:supp-selected-primary}
\resizebox{\textwidth}{!}{%
\begin{tabular}{lcccc ccccc}
\toprule
Scenario & \multicolumn{3}{c}{BOIN-CS} & \multicolumn{3}{c}{BOIN-CB} & \multicolumn{3}{c}{BOIN-CE} \\
\cmidrule(lr){2-4} \cmidrule(lr){5-7} \cmidrule(lr){8-10}
 & Correct\% & Acceptable\% & Over\% & Correct\% & Acceptable\% & Over\% & Correct\% & Acceptable\% & Over\% \\
\midrule
\multicolumn{10}{c}{Configuration $\mathcal{C}_s$}\\
\midrule
1 & 32.0 & 74.6 & 0.0 & 31.4 & 74.7 & 0.0 & 30.2 & 77.1 & 0.0 \\
2 & 44.5 & 67.0 & 13.8 & 41.6 & 64.9 & 15.4 & 45.4 & 68.5 & 12.4 \\
3 & 22.7 & 64.4 & 14.6 & 7.6 & 61.1 & 19.2 & 24.8 & 64.9 & 14.1 \\
4 & 22.3 & 72.0 & 10.4 & 32.5 & 79.6 & 4.1 & 22.8 & 73.3 & 9.0 \\
5 & 65.4 & 65.4 & 5.2 & 65.2 & 65.2 & 4.6 & 65.3 & 65.3 & 5.4 \\
6 & 50.6 & 50.6 & 19.6 & 51.4 & 51.4 & 19.9 & 50.6 & 50.6 & 19.6 \\
7 & 0.0 & 52.2 & 29.4 & 0.0 & 53.5 & 28.1 & 0.0 & 52.2 & 29.1 \\
8 & 24.7 & 72.7 & 13.7 & 24.7 & 72.7 & 13.7 & 25.4 & 73.9 & 12.5 \\
9 & 0.0 & 39.9 & 50.5 & 0.0 & 42.3 & 49.0 & 0.0 & 39.9 & 50.5 \\
10 & 52.8 & 52.8 & 19.3 & 53.9 & 53.9 & 19.6 & 52.8 & 52.8 & 19.3 \\
11 & 0.0 & 62.8 & 28.0 & 0.0 & 64.7 & 27.9 & 0.0 & 62.8 & 28.0 \\
12 & 51.4 & 85.6 & 11.0 & 51.4 & 84.7 & 13.0 & 51.4 & 85.6 & 11.0 \\
13 & 56.0 & 56.0 & 29.3 & 58.6 & 58.6 & 28.6 & 56.0 & 56.0 & 29.3 \\
14 & 0.0 & 0.0 & 59.0 & 0.0 & 0.0 & 58.7 & 0.0 & 0.0 & 59.0 \\
\midrule
Mean & 30.2 & 58.3 & 21.7 & 29.9 & 59.1 & 21.6 & 30.3 & 58.8 & 21.4 \\
\bottomrule
\multicolumn{10}{c}{Configuration $\mathcal{C}_{Full}$}\\
\midrule
1 & 33.7 & 80.6 & 0.0 & 36.2 & 82.1 & 0.0 & 33.4 & 78.4 & 0.0 \\
2 & 40.3 & 65.9 & 15.9 & 44.2 & 69.6 & 14.0 & 40.5 & 69.2 & 13.2 \\
3 & 21.6 & 70.7 & 15.1 & 12.8 & 68.0 & 18.0 & 20.2 & 69.7 & 15.2 \\
4 & 17.6 & 64.5 & 19.0 & 29.8 & 74.9 & 10.8 & 15.5 & 65.5 & 18.1 \\
5 & 29.2 & 65.2 & 19.0 & 34.4 & 76.3 & 11.6 & 25.2 & 59.2 & 21.4 \\
6 & 55.1 & 55.1 & 26.7 & 54.8 & 54.8 & 26.2 & 54.5 & 54.5 & 24.5 \\
7 & 28.8 & 54.4 & 28.3 & 20.2 & 52.4 & 32.6 & 23.6 & 54.3 & 27.2 \\
8 & 31.4 & 64.1 & 19.8 & 35.3 & 72.3 & 13.1 & 31.3 & 63.3 & 17.6 \\
9 & 26.0 & 57.4 & 37.4 & 22.8 & 52.7 & 43.1 & 26.5 & 60.7 & 34.9 \\
10 & 56.1 & 56.1 & 25.6 & 56.3 & 56.3 & 28.0 & 58.1 & 58.1 & 25.3 \\
11 & 15.1 & 41.1 & 40.1 & 9.9 & 40.2 & 40.1 & 19.5 & 46.7 & 36.5 \\
12 & 25.4 & 53.7 & 43.6 & 27.8 & 55.5 & 41.7 & 27.5 & 55.4 & 42.2 \\
13 & 52.8 & 52.8 & 34.9 & 53.6 & 53.6 & 35.6 & 50.1 & 50.1 & 35.2 \\
14 & 0.0 & 0.0 & 57.5 & 0.0 & 0.0 & 58.5 & 0.0 & 0.0 & 58.3 \\
\midrule
Mean & 30.9 & 55.8 & 27.3 & 31.3 & 57.8 & 26.7 & 30.4 & 56.1 & 26.4 \\
\bottomrule
\end{tabular}%
}
\end{table}

\begin{table}[H]
\centering
\caption{Operating Characteristics: Average number of patients treated at overly toxic combinations (PtsOverTox), Total number of DLTs (TotalDLT), Total Realized Sample Size (TotalN)}
\label{tab:supp-selected-secondary}
\resizebox{\textwidth}{!}{%
\begin{tabular}{lccccccccc}
\toprule
Scenario & \multicolumn{3}{c}{BOIN-CS} & \multicolumn{3}{c}{BOIN-CB} & \multicolumn{3}{c}{BOIN-CE} \\
\cmidrule(lr){2-4} \cmidrule(lr){5-7} \cmidrule(lr){8-10}
 & PtsOverTox & TotalDLT & TotalN & PtsOverTox & TotalDLT & TotalN & PtsOverTox & TotalDLT & TotalN \\
\midrule
\multicolumn{10}{c}{Configuration $\mathcal{C}_s$}\\
\midrule
1 & 0.0 & 5.6 & 32.1 & 0.0 & 5.6 & 32.2 & 0.0 & 5.7 & 32.4 \\
2 & 3.3 & 6.5 & 33.2 & 3.2 & 6.5 & 33.4 & 3.1 & 6.7 & 34.2 \\
3 & 3.7 & 6.3 & 33.0 & 3.8 & 6.3 & 33.2 & 3.5 & 6.4 & 33.4 \\
4 & 3.6 & 6.8 & 31.4 & 2.3 & 6.6 & 31.3 & 3.9 & 7.1 & 32.0 \\
5 & 5.0 & 7.7 & 32.4 & 4.9 & 7.8 & 32.7 & 4.9 & 7.6 & 32.3 \\
6 & 5.5 & 6.6 & 25.0 & 5.5 & 6.6 & 25.2 & 5.5 & 6.6 & 25.0 \\
7 & 7.7 & 6.5 & 25.9 & 7.8 & 6.6 & 26.0 & 7.6 & 6.5 & 25.9 \\
8 & 4.2 & 6.8 & 32.0 & 4.2 & 6.8 & 32.0 & 4.3 & 7.1 & 32.7 \\
9 & 9.9 & 6.3 & 23.4 & 9.8 & 6.3 & 23.4 & 9.9 & 6.3 & 23.4 \\
10 & 5.0 & 6.1 & 20.8 & 5.6 & 6.3 & 21.4 & 5.0 & 6.1 & 20.8 \\
11 & 7.6 & 6.4 & 26.0 & 7.3 & 6.2 & 25.6 & 7.7 & 6.4 & 26.0 \\
12 & 4.1 & 5.9 & 19.0 & 4.3 & 6.0 & 19.3 & 4.1 & 5.9 & 19.0 \\
13 & 6.7 & 5.4 & 15.2 & 6.7 & 5.4 & 15.1 & 6.7 & 5.4 & 15.2 \\
14 & 10.1 & 4.7 & 10.1 & 10.2 & 4.7 & 10.1 & 10.1 & 4.7 & 10.1 \\
\midrule
Mean & 5.5 & 6.3 & 25.7 & 5.4 & 6.3 & 25.8 & 5.5 & 6.3 & 25.9 \\
\bottomrule
\multicolumn{10}{c}{Configuration $\mathcal{C}_{Full}$}\\
\midrule
1 & 0.0 & 5.7 & 32.9 & 0.0 & 5.7 & 33.2 & 0.0 & 5.7 & 33.5 \\
2 & 3.3 & 6.5 & 33.5 & 3.4 & 6.5 & 33.6 & 3.1 & 6.8 & 35.3 \\
3 & 3.6 & 6.4 & 33.4 & 3.9 & 6.4 & 33.9 & 3.4 & 6.6 & 35.1 \\
4 & 5.3 & 7.2 & 32.5 & 3.7 & 7.2 & 33.3 & 5.7 & 7.5 & 34.2 \\
5 & 7.0 & 7.9 & 33.1 & 6.1 & 7.9 & 33.8 & 7.8 & 8.4 & 35.4 \\
6 & 7.1 & 8.0 & 29.2 & 7.8 & 8.4 & 29.8 & 7.0 & 8.4 & 30.9 \\
7 & 7.5 & 7.6 & 29.8 & 9.4 & 8.5 & 31.8 & 8.1 & 8.3 & 32.1 \\
8 & 5.5 & 7.2 & 33.0 & 4.3 & 7.3 & 33.9 & 6.0 & 7.8 & 35.0 \\
9 & 9.1 & 7.7 & 27.4 & 11.3 & 8.5 & 29.3 & 9.1 & 8.1 & 29.1 \\
10 & 7.4 & 7.8 & 25.3 & 9.0 & 8.5 & 26.8 & 7.8 & 8.3 & 26.9 \\
11 & 10.8 & 8.1 & 30.7 & 11.6 & 8.6 & 32.1 & 11.2 & 8.9 & 33.9 \\
12 & 10.7 & 7.7 & 23.5 & 11.1 & 8.2 & 24.3 & 10.9 & 8.0 & 24.5 \\
13 & 8.6 & 6.4 & 17.3 & 9.5 & 6.8 & 18.3 & 9.5 & 6.9 & 18.5 \\
14 & 10.9 & 5.1 & 10.9 & 11.0 & 5.2 & 11.1 & 11.0 & 5.1 & 11.0 \\
\midrule
Mean & 6.9 & 7.1 & 28.0 & 7.3 & 7.4 & 28.9 & 7.2 & 7.5 & 29.7 \\
\bottomrule
\end{tabular}%
}
\end{table}

\bibliography{boin-ref}
\bibliographystyle{plainnat}